\documentclass[aps,prl,reprint,superscriptaddress]{revtex4-1}
\usepackage{graphicx}
\usepackage[rightcaption]{sidecap}
\usepackage{braket}
\usepackage[utf8]{inputenc}
\usepackage{amsmath} 
\usepackage{amssymb}  
\usepackage{amstext}  
\graphicspath{{./figures/}}

\begin{document}
\title{Coherence of a dynamically decoupled quantum-dot hole spin }

\author{L. Huthmacher}
\affiliation{Cavendish Laboratory, University of Cambridge, JJ Thomson Avenue, Cambridge, CB3 0HE, UK}

\author{R. Stockill}
\affiliation{Cavendish Laboratory, University of Cambridge, JJ Thomson Avenue, Cambridge, CB3 0HE, UK}

\author{E. Clarke}
\affiliation{EPSRC National Centre for III-V Technologies, University of Sheffield, Sheffield, S1 3JD, UK}

\author{M. Hugues}
\affiliation{Universit\'e C\^ote d'Azur, CNRS, CRHEA, Valbonne, France}

\author{C. Le Gall}
\affiliation{Cavendish Laboratory, University of Cambridge, JJ Thomson Avenue, Cambridge, CB3 0HE, UK}

\author{M. Atat{\"u}re}
\email[Electronic address: ]{ma424@cam.ac.uk}
\affiliation{Cavendish Laboratory, University of Cambridge, JJ Thomson Avenue, Cambridge, CB3 0HE, UK}

\begin{abstract}
A heavy hole confined to an InGaAs quantum dot promises the union of a stable spin and optical coherence to form a near perfect, high-bandwidth spin-photon interface. Despite theoretical predictions and encouraging preliminary measurements, the dynamic processes determining the coherence of the hole spin are yet to be understood. Here, we establish the regimes that allow for a highly coherent hole spin in these systems, recovering a crossover from hyperfine to electrical-noise dominated decoherence with a few-Tesla external magnetic field. Dynamic decoupling allows us to reach the longest ground-state coherence time, $T_2$, of $4.4\, \mu$s, observed in this system. The improvement of coherence we measure is quantitatively supported by an independent analysis of the local electrical environment.
\end{abstract}

\pacs{}

\date{\today}

\maketitle

\indent Self-assembled indium gallium arsenide (InGaAs) quantum dots (QDs) provide an excellent test bed for tackling the implementation challenges of distributed quantum information processing \cite{Kimble2008a,Gao2015a}. They can be charged deterministically \cite{Smith2005} and both electrons and holes can serve as optically active qubits. The outstanding photonic properties \cite{Ding2016} in combination with ultrafast spin control \cite{Press2008b,DeGreve2011} and efficient state transfer through spin-photon entanglement \cite{DeGreve2012b,Gao2012a,Schaibley2013} have recently allowed for the generation of spin-spin entanglement \cite{Delteil2015,Stockill2017}.
The first key figure of merit of such a system is the inhomogeneous dephasing time $T_2^*$ relative to the longest operation time, which, in the case of QDs, is the spin-photon state transfer characterized by the optical lifetime $\Gamma^{-1}$. The second figure of merit is the spin coherence time $T_2$, which determines for how long the entanglement can be preserved.
For the electron spin, $T_2^*$ of a few ns (comparable to $\Gamma^{-1},\, \approx 0.7\,\mathrm{ns}$) and $T_2$ of a few $\mu$s \cite{Bechtold2015a,Stockill2016} are predominantly limited by the size and dispersion of the QD nuclear spin ensemble, respectively.
For the hole spin, the dominant contact interaction term is suppressed due to hole's \textit{p}-like symmetry, offering an order of magnitude weaker hyperfine interaction \cite{Fischer2008,Fallahi2010}.
Additionally, given the predominantly heavy-hole character of the ground state, the hyperfine interaction is primarily concentrated along the growth axis and can be suppressed by a transverse external magnetic field \cite{Fischer2008}.\\
\indent The coherence of the hole spin has been observed in a number of experiments: Studies using coherent population trapping (CPT) have suggested a promising $T_2^* \geq 100\, \mathrm{ns}$ \cite{Houel2014, Prechtel2016}, whereas direct measurements of the free induction decay through Ramsey interference \cite{Greilich2011,DeGreve2011,Godden2012} and spin-flip Raman scattering \cite{Sun2016} could only reach $T_2^*$ of up to $26\,\mathrm{ns} $. 
While nuclear-spin noise and electrical charge fluctuations have both been suggested as the dominating source of decoherence for the hole spin \cite{DeGreve2011, Godden2012, Houel2014, Prechtel2016}, the understanding of the mechanism governing the coherence and, more importantly, how well it can be protected remains unclear.
In this Letter, we study the performance of a single hole spin in an InGaAs QD experiencing a dynamic nuclear and electric environment. The dependence of $T_2^*$ and $T_2$ on the external magnetic field reveals significant coupling to the nuclear spin ensemble at low fields and to electrical noise at high fields. Our results indicate that strain-induced mixing with the light-hole states enables hyperfine interactions that bound the coherence time for external magnetic fields up to a few Tesla. 
At higher fields we prolong the hole spin coherence by employing a dynamic decoupling sequence with an increase of coherence time solely determined by the underlying electrical noise spectrum.\\
\indent The self-assembled InGaAs QDs are embedded in a n-type Schottky diode heterostructure \cite{Supp}, which is cooled to $4.2\,\mathrm{K}$. We drive the neutral exciton transition resonantly around 970 nm under a constant DC bias such that the electron tunnels out nearly instantaneously, leaving behind a hole with a charge lifetime exceeding $45 \, \mu\mathrm{s}$.
The external magnetic field $B_{\mathrm{ext}}$ perpendicular to the growth axis lifts the degeneracy of the ground and excited states with a ground-state splitting of $ \approx2.3\,\mathrm{GHz}\mathrm{T}^{-1}$. The spin state is initialized and read out by driving one of the four transitions of the positively charged trion around 971 nm and coherently manipulated using far red-detuned picosecond laser pulses \cite{DeGreve2011}.
\\
\begin{figure}
\includegraphics[width = 1\columnwidth,angle=0]{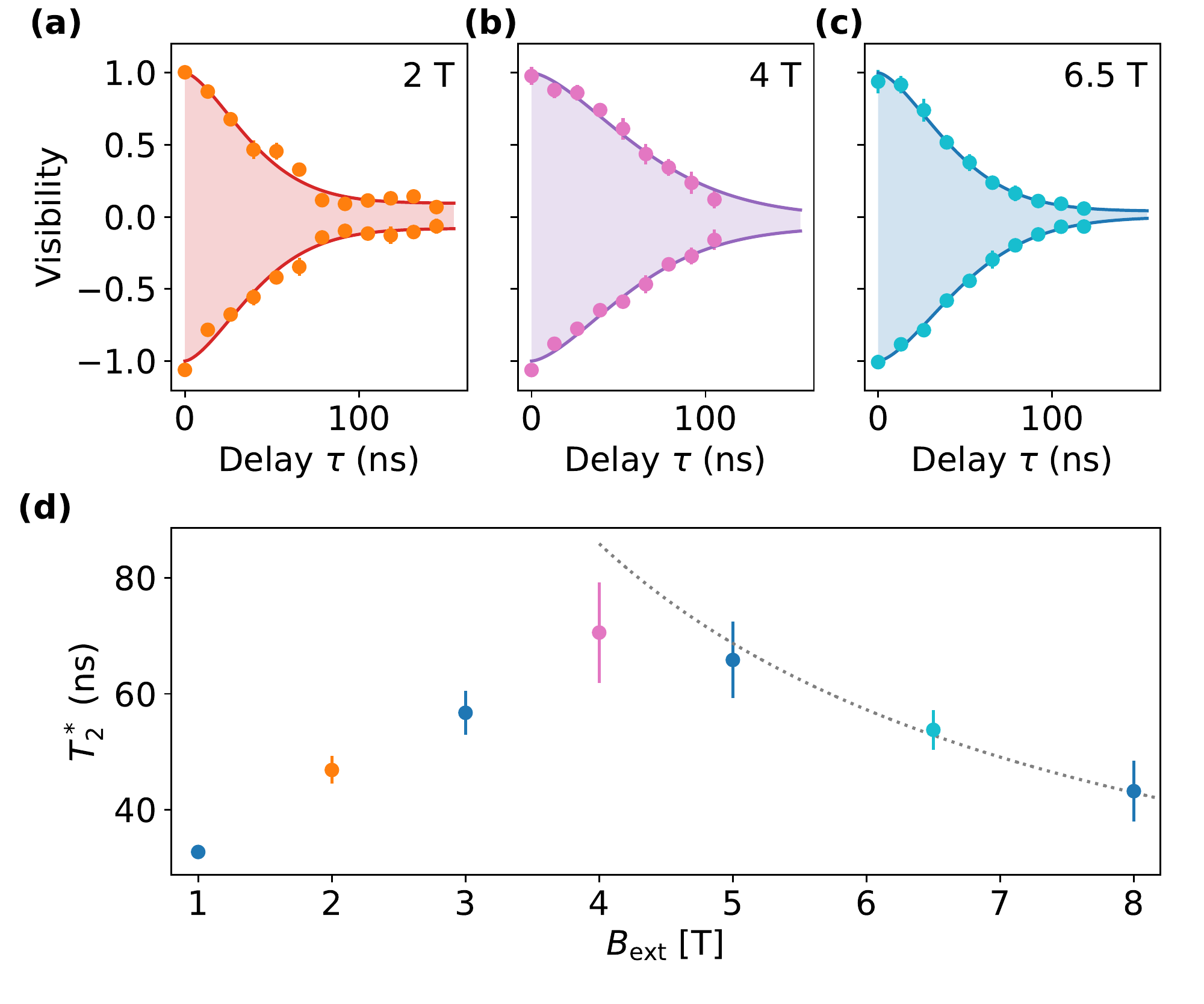}
\caption{Measurement of magnetic field dependent inhomogeneous dephasing time, $T_2^*$. (a)-(c) Visibility of Ramsey fringes measured at external magnetic fields $B_{\mathrm{ext}}$ of  $2\,\mathrm{T}$ (orange), $4\,\mathrm{T}$ (purple), $6.5\,\mathrm{T}$ (light blue). Error bars represent $\pm$ 1 standard deviation. Solid curves are fits to the data to extract $T_2^*$. (d) Summary of magnetic field dependent measurement of $T_2^*$, data points from panels (a)-(c) are presented in the corresponding color and error bars represent $\pm$ 1 standard deviation. Grey dotted curve shows decay $\propto 1/B_\mathrm{ext}$ for $B_{\mathrm{ext}}>4\,T$.}
\label{fig:1}
\end{figure}
\indent In order to assess the quality of the hole-based spin-photon interface we first study the inhomogeneous dephasing time $T_2^*$. In Fig. \ref{fig:1} (a-c) we present the decay envelope of the Ramsey-interference visibility \cite{Press2008b} measured at three different external magnetic fields. The inhomogeneous dephasing times extracted from fits to the data are presented in Fig. \ref{fig:1} (d) for the full range of $1\,\mathrm{T}\leq B_{\mathrm{ext}} \leq 8\,\mathrm{T}$. These dephasing times are an order of magnitude larger than the ones observed for electrons \cite{Bechtold2015a,Stockill2016}, with $T_2^*$ peaking at $70.6 \pm 8.7 \, \mathrm{ns}$ for $B_{\mathrm{ext}}=4\,\mathrm{T}$. In the high field regime ($B_{\mathrm{ext}} > 4 \, \mathrm{T}$) we observe a decay of $T_2^*$, which is proportional to $1/B_{\mathrm{ext}}$, indicated by the grey dotted curve in Fig. \ref{fig:1} (d). The behavior is a clear evidence of electrical noise induced inhomogeneous dephasing as suggested by Houel \textit{et al.} \cite{Houel2014}: The local electric field $F$ influences the exact position of the hole wave function within the QD. This in turn determines the in-plane hole g factor $\frac{2}{\sqrt{3}}\beta g_\mathrm{h}$ \cite{Salis2001}, where $\beta$ is the light-hole component of a predominantly heavy-hole state. Consequently, electrical noise $\delta F$ in the sample affects the ground-state Zeeman splitting following a linear magnetic field dependence $\delta E_Z^{\mathrm{elec}}=\left( \frac{2}{\sqrt{3}}\frac{\partial (\beta g_\mathrm{h})}{\partial F} \right) \delta F \mu_\mathrm{B} B_{\mathrm{ext}}$, leading to $T_2^* \propto 1/(\delta F B_{\mathrm{ext}})$.\\
\indent We find that in the low field regime $T_2^*$ displays a linear dependence on magnetic field. This evolution of the coherence was predicted to occur for heavy holes with negligible light-hole admixture \cite{Fischer2008,Testelin2009} as a result of the hyperfine coupling to the nuclear spin ensemble. 
The nuclear spin fluctuations $\delta {B}^{\mathrm{nuc}}$ affect the ground-state splitting by $\delta E_Z^{\mathrm{nuc}}$ according to:

\begin{equation}
\delta E_Z^{\mathrm{nuc}} \approx \mu_\mathrm{B} g_\mathrm{h} \delta B_x^{\mathrm{nuc}} + \frac{{\mu_\mathrm{B} g_\mathrm{h}}}{2\beta B_{\mathrm{ext}}}{(\delta B_z^{\mathrm{nuc}})^2}.
\end{equation}

\noindent Here, $\delta B_x^{\mathrm{nuc}}$  and $\delta B_z^{\mathrm{nuc}}$ are the effective fields arising from nuclear spin fluctuations along the external magnetic field and the growth axis, which affect the ground-state splitting to first and second order, respectively. 
For this QD, we estimate that $\delta B_z^{\mathrm{nuc}}\approx 0.8\,\mathrm{mT}$ from the dephasing of the electron spin \cite{Supp}. We infer a light-hole component of $\beta\approx 0.08$ from the Zeeman splitting, leading to $\delta B_x^{\mathrm{nuc}}\approx0.07\,\mathrm{mT}$. These values are consistent with estimates inferred from hole depolarization \cite{Eble2009, Dahbashi2012}. 
The linear increase of $T_2^*$ with $B_{\mathrm{ext}}$ should be observed if the second-order term dominates, i.e. $\frac{\delta B_z^{\mathrm{nuc}}}{2\beta^2B_{\mathrm{ext}}}\gg1$. For the QD studied here, our estimates indicate that we are not in this regime if we assume an unperturbed nuclear spin ensemble and the linear increase of dephasing time could arise from back-action on the nuclei.
The hole-spin $T_2^*$ values we extract indicate a spin-photon interface superior to the electron over a large range of external magnetic fields. By comparing the dephasing time to the optical recombination time $\Gamma^{-1} \approx 0.7$ ns, the electron $T_2^*$ of 2.2 ns measured in the same QD \cite{Supp} bounds the fidelity of an entangled spin-photon state to $92\%$. In contrast, the ground-state dephasing of the hole only leads to a loss of $< 0.1\%$ fidelity for all values of $B_{\mathrm{ext}}$ shown here, an improvement of almost 2 orders of magnitude.\\
\begin{figure}
\includegraphics[width = 1\columnwidth,angle=0]{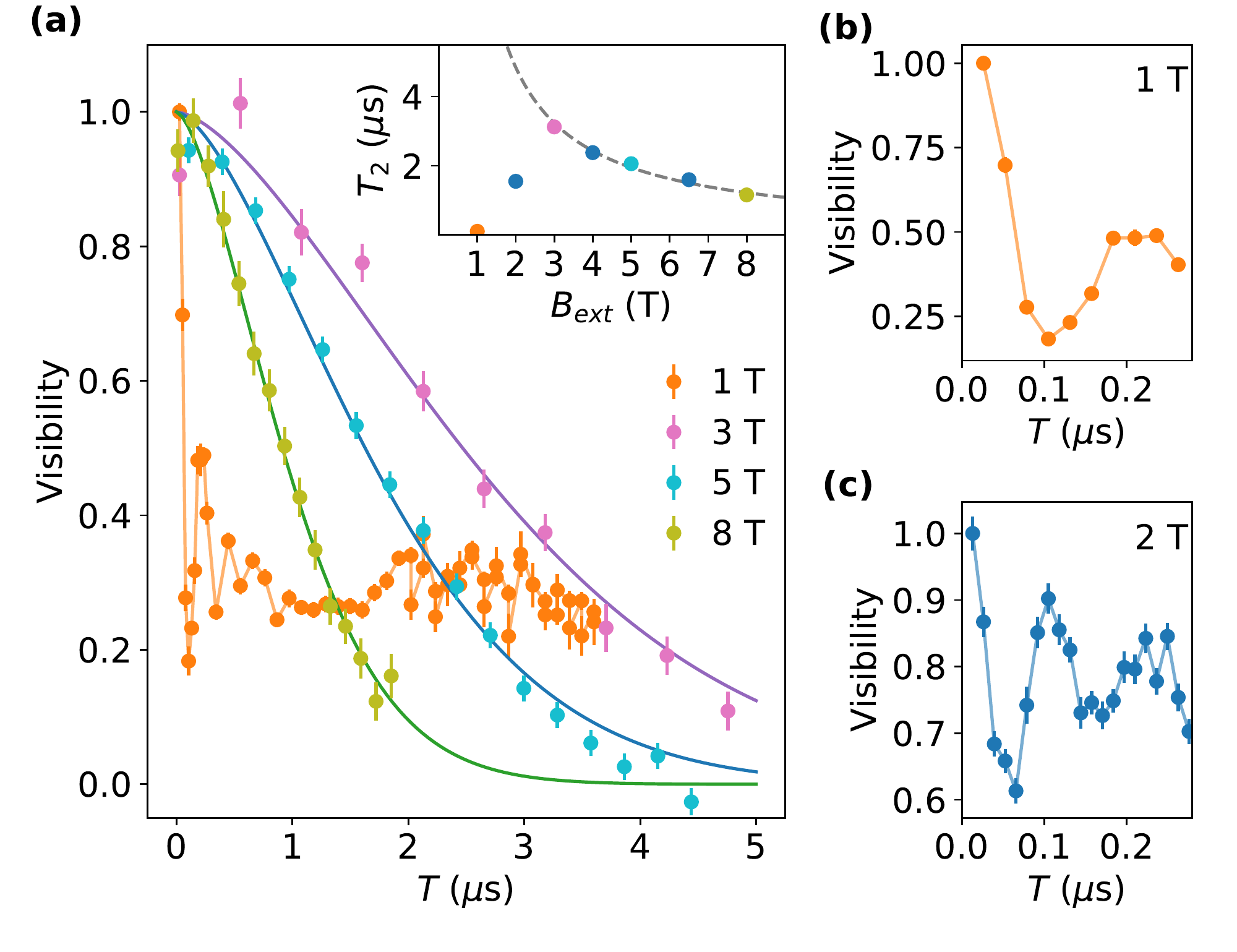}
\caption{Hahn-echo measurement for different values of $B_{\mathrm{ext}}$. (a) Visibility of the Hahn-echo signal for $B_{\mathrm{ext}}$ of 1 T (orange), 3 T (purple), 5 T (light blue) and 8 T (green). Error bars represent $\pm$ 1 standard deviation. Solid curves are fits to extract $T_2$, for $B_{\mathrm{ext}}=1\,\mathrm{T}$ the solid curve only serves as guide to the eye. The inset shows the full behavior of $T_2$ with respect to $B_{\mathrm{ext}}$, values are extracted from the fits for $B_{\mathrm{ext}} >2\,\mathrm{T}$ and error bars represent $\pm$ 1 standard deviation. For $B_{\mathrm{ext}}\leq 2\,\mathrm{T}$ we show the time where the visibility falls below $1/e$ for the first time. (b,c) Zoom in for the 1- and 2-T data, revealing a sharp drop and revival of coherence within the first 300 ns.}
\label{fig:2}
\end{figure}
\indent We study the extent to which the quantum state of a hole spin can be preserved through decoupling techniques, namely the coherence time $T_2$. Adding a refocusing pulse in the center of the Ramsey sequence implements a Hahn-echo measurement \cite{Hahn1950}, which suppresses the effect of correlated noise on the system. The results are presented in Fig. \ref{fig:2} (a), where we show four example measurements at different magnetic fields together with the corresponding fits of $V(\tau)=V_0  \exp [-(\tau/T_2)^{1.56}]$ \footnote{The choice of $\alpha$ is motivated by the scaling observed in dynamic decoupling \cite{Supp}}. The magnetic field dependence of $T_2$ is shown in the inset. For fields $\geq 3 \,\mathrm{T}$ we observe a decrease of the coherence time with increasing external magnetic field $T_2(B_\mathrm{ext}) \propto 1/B_\mathrm{ext}^{0.99 \pm0.03}$ which can be understood, similarly to the case of $T_2^*$, by considering the increase in coupling to electric noise. While the loss of coherence at higher fields is approximated by a single decay, we observe structure in the Hahn-echo visibility for low external magnetic fields. Panels (b) and (c) of Fig. \ref{fig:2} show the initial 280 ns for the two lowest measured values of $B_{\mathrm{ext}}$. In both cases the data show a sharp decay and revival of visibility. This behavior has also been observed for the electron and is due to interactions with the precessing nuclei \cite{Bechtold2015a,Stockill2016}. Specifically, the first-order coupling leads to spectral features at nuclear Zeeman-splitting frequencies which result in modulation of the echo signal at short delays and low fields \cite{Supp}. Even though the pseudospin studied here has a strong heavy-hole character, the effect of nuclear fluctuations along the external field still clearly dominates the hole dynamics. \\
\begin{figure}
\includegraphics[width = 1\columnwidth,angle=0]{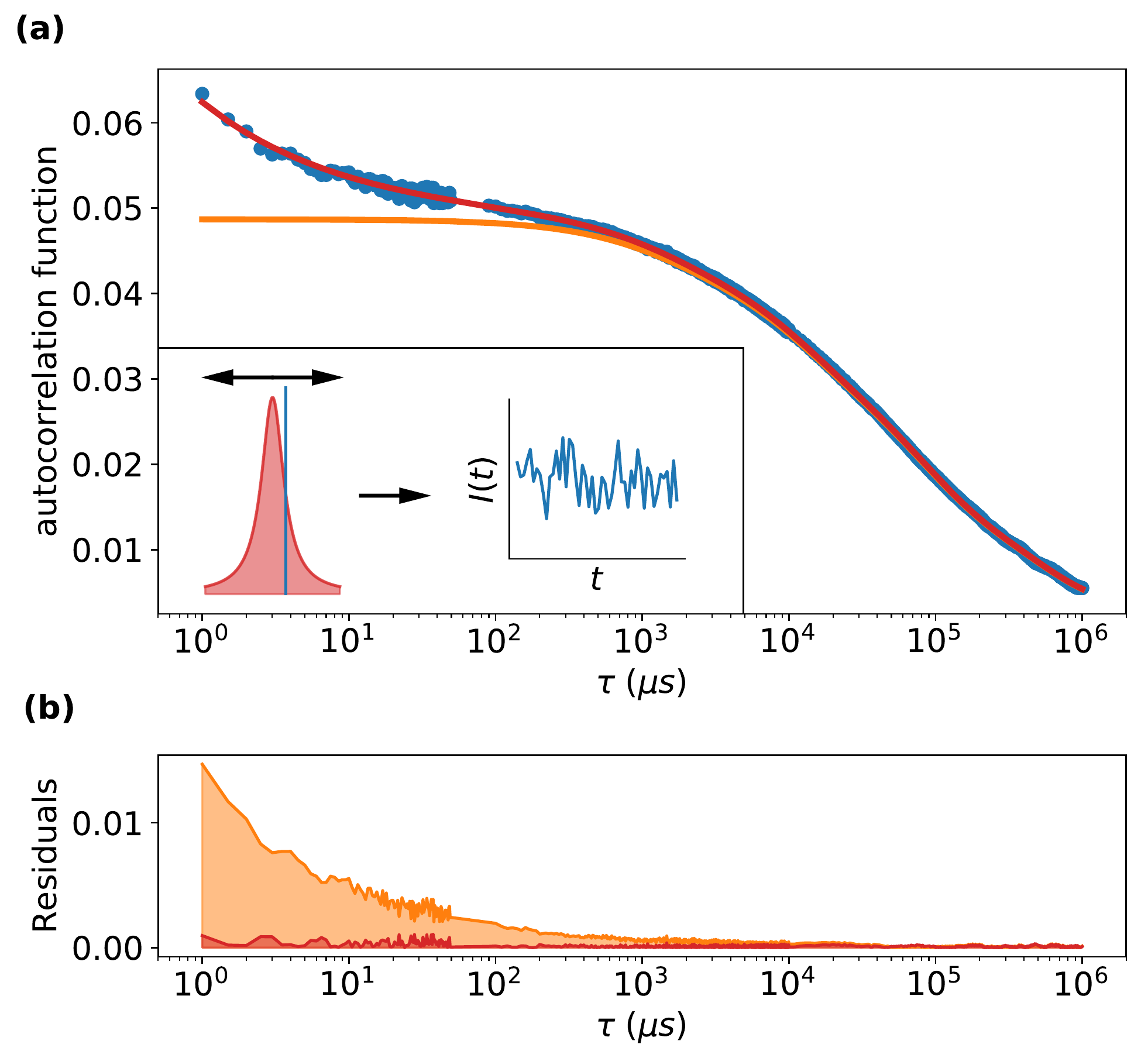}
\caption{Normalized autocorrelation function of the neutral exciton intensity fluctuations. (a) Extracted autocorrelation function of intensity fluctuations measured on the neutral exciton transition (solid blue circles). The red curve represents a fit to the data containing 4 exponential functions as well as a $1/\tau^{1-\lambda}$ component. The orange curve represents only the contribution of the exponential functions. The inset is an illustration of how electrical noise leads to intensity fluctuations, responsible for the bunching of the autocorrelation function. (b) Residuals of the two curves presented in (a), highlighting the strong deviation of the orange curve from the data for $\tau < 1\, \mathrm{ms}$.}
\label{fig:3} 
\end{figure}
\indent Electrical noise also leads to shifts of the QD optical frequency via the Stark effect, detuning it from the resonant laser drive, with the effect of changing the intensity of scattered light, $I(t)$, over time. Calculating the Fourier transform \cite{Kuhlmann2013a} or equivalently the normalised autocorrelation function $\left(\left\langle I\left(\tau\right)I\left(0\right)\right\rangle/\left\langle I\right\rangle^2 -1\right)$ \cite{Stanley2014} of this signal allows us to directly access the properties of electrical noise in the environment of the QD. Figure \ref{fig:3} (a) presents the autocorrelation of resonantly scattered light from the neutral exciton transition of the QD studied in this work. The main features in the autocorrelation data are attributed to two-level fluctuators, which result in exponential decays of different time scales (orange curve). Those noise sources contribute significantly to the hole $T_2^*$, but are suppressed efficiently using the Hahn-echo technique, due to their slow switching time ($\gtrsim$ 1 ms). Additionally, the data reveal a $1/\tau$-like component (included in the red curve), particularly apparent at small values of $\tau$ [Fig. 3(b)]. This can be related to a $1/f^{\lambda}$ noise spectrum which results in an autocorrelation function of the form $1/\tau^{1-\lambda}$ for $\lambda < 1$ \cite{Keshner1982}. Fitting the data with the combination of exponential decays and a $1/\tau^{1-\lambda}$ function, we extract $\lambda = 0.56\pm 0.01$. The high-frequency tail of this noise, for which the exact origin remains unclear \cite{Dial2013,Kuhlmann2013a,Houel2014}, limits the efficacy of Hahn echo, and thus the coherence of the hole spin. Scaling the coupling of the $1/f^{\lambda}$ component to fit the Hahn-echo decay, we can infer the contribution of the low-frequency noise to the ensemble dephasing time, $T_2^*$ \cite{Supp}. For example at an external field of $B_{\mathrm{ext}} = 6.5$ T, we find that the autocorrelation data predict $T_2^* = 55$ ns, is consistent with our measured value of $T_2^* = 53.8 \pm 3.4$ ns.
In fact, knowledge of the electrical noise and its coupling enables us to capture the high-field coherent dynamics of the hole spin in its entirety.\\
\begin{figure*}
\includegraphics[width=0.99\textwidth]{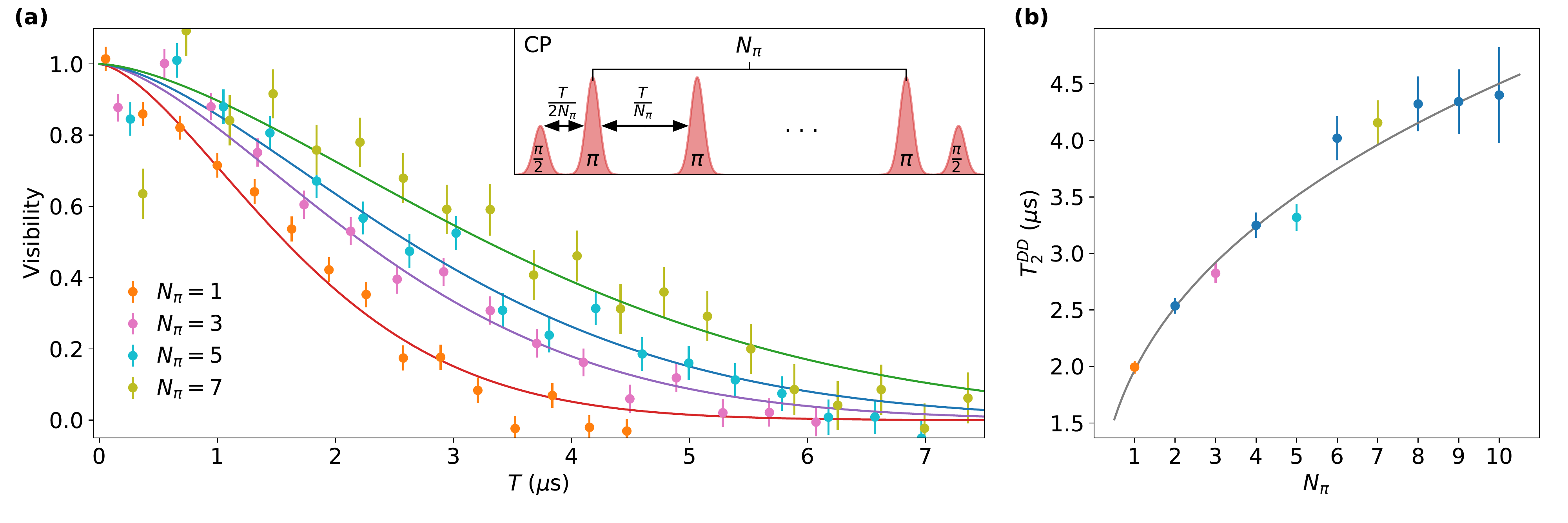}
\caption{Dynamic decoupling of the hole spin. (a) Visibility for dynamic decoupling at $B_{\mathrm{ext}}=5\,\mathrm{T}$ as a function of the number of $\pi$ pulses, where $N_{\pi}$ is 1 (orange), 3 (purple), 5 (light blue) and 7 (green). The inset shows a schematic of the employed pulse sequence. Solid curves represent fits to the data to extract the coherence time and error bars are given by $\pm$ 1 standard deviation. (b) Scaling of the coherence time with the number of $\pi$ pulses, data presented in (a) shown in matching color. Grey curve presents a fit of $T_2^0 (N_\pi)^\gamma$ to the data, extracted scaling $\gamma = 0.36 \pm 0.02$. Error bars represent $\pm$ 1 standard deviation.}
\label{fig:4}
\end{figure*}
\indent We can protect the hole actively against high-frequency electrical noise and explore the limit to hole spin coherence by implementing a dynamic decoupling scheme. The data in Fig. \ref{fig:2} indicate that the electrical noise dominates in the regime of $B_{\mathrm{ext}}\geq 3\,\mathrm{T}$ and we thus choose to work at $B_{\mathrm{ext}}=5\,\mathrm{T}$. The sequence we implement is based on the first proposal by Carr and Purcell (CP) \cite{CarrH.Y.1983}, shown in the inset of Fig. \ref{fig:4} (a), where multiple refocusing pulses are concatenated. Figure \ref{fig:4} (a) shows the visibility decay for different number of refocusing pulses $N_\pi =1,3,5,7$. As an extension of the simple Hahn echo sequence this decay is also described by $V(\tau)=V_0  \exp [-(\tau/T_2)^\alpha]$. In order to determine the value of $\alpha$ we follow the approach of J. Medford \textit{et al.} \cite{Medford2012} and extract $\alpha = 1.56 \pm 0.01$ from the scaling of coherence with $N_\pi$ \cite{Supp}. As expected, we do observe a pick up of $T_2^{DD}$ with increasing number of refocusing pulses (see Fig. \ref{fig:4} (b)), which is described by $T_2^{DD}\left(N_\pi\right) = T_2 (N_\pi)^\gamma$ with $\gamma = 0.36 \pm 0.02$.  
We successfully decouple the hole from the electrical noise, preserve the coherence and reach dephasing times higher than for electrons, in this work $T_2^{DD} = 4.40\pm0.43\, \mu\mathrm{s}$ for $N_\pi = 10$. The main limitation for applications is the fidelity of the refocusing pulses. We employ composite pulses to achieve higher spin rotation fidelities \cite{Levit1986}, but ultimately the visibility decreases with increasing $N_\pi$ \cite{Supp}. Extracting the scaling with $N_\pi$ also allows us to check that we are efficently decoupling the system from the underlying noise source. In the case of $1/f$ noise the scaling is directly linked to the exponent of the noise spectrum through  $\lambda=\gamma/\left(1 - \gamma\right)$ \cite{Medford2012}. Therefore, we extract $\lambda=0.56$ which agrees with the value extracted from the autocorrelation function of the electrical noise in Fig. \ref{fig:3}, showing that we decouple our system. \\
\indent In this work we have observed how a hole spin interacts with a dynamic solid-state environment. We reveal a crossover between low- and high-field regimes in both the $T_2^*$ and the $T_2$. We report the longest coherence times for a freely evolving $\left(T_2^*\right)$ and protected $\left(T_2^{DD}\right)$ hole spin in direct transient measurements. Surprisingly, given the weak hyperfine coupling of the heavy hole and the relatively small light-hole hybridization of the studied QD, we find that the nuclear environment still dictates the coherence time up to fields of a few Tesla. The linear low-field dependence we observe in $T_2^*$, while appearing to follow a second-order coupling to nuclei, results in deviations from previously measured values for the hole-hyperfine coupling constant. At high fields we find that the coherence time of the spin is limited by the electric field dependence of the g-factor. We show that the underlying noise spectrum, $\sim 1/f^{0.6}$, sets the scaling of the extension of coherence to $N_\pi^0.4$. The limiting noise source identified here is extrinsic to the QD and an understanding of the exact origin of his $1/f^{\lambda}$ noise is key to further prolongation of the coherence time.\\

\indent We gratefully acknowledge financial support by the European Research Council ERC Consolidator grant agreement no. 617985 and the EPSRC
National Quantum Technologies Programme NQIT EP/M013243/1. We thank M.J. Stanley, G. \' Ethier-Majcher, D. Gangloff, J. Bodey and C. Lang for fruitful discussions.

%

\begin{thebibliography}{34}%
\makeatletter
\providecommand \@ifxundefined [1]{%
 \@ifx{#1\undefined}
}%
\providecommand \@ifnum [1]{%
 \ifnum #1\expandafter \@firstoftwo
 \else \expandafter \@secondoftwo
 \fi
}%
\providecommand \@ifx [1]{%
 \ifx #1\expandafter \@firstoftwo
 \else \expandafter \@secondoftwo
 \fi
}%
\providecommand \natexlab [1]{#1}%
\providecommand \enquote  [1]{``#1''}%
\providecommand \bibnamefont  [1]{#1}%
\providecommand \bibfnamefont [1]{#1}%
\providecommand \citenamefont [1]{#1}%
\providecommand \href@noop [0]{\@secondoftwo}%
\providecommand \href [0]{\begingroup \@sanitize@url \@href}%
\providecommand \@href[1]{\@@startlink{#1}\@@href}%
\providecommand \@@href[1]{\endgroup#1\@@endlink}%
\providecommand \@sanitize@url [0]{\catcode `\\12\catcode `\$12\catcode
  `\&12\catcode `\#12\catcode `\^12\catcode `\_12\catcode `\%12\relax}%
\providecommand \@@startlink[1]{}%
\providecommand \@@endlink[0]{}%
\providecommand \url  [0]{\begingroup\@sanitize@url \@url }%
\providecommand \@url [1]{\endgroup\@href {#1}{\urlprefix }}%
\providecommand \urlprefix  [0]{URL }%
\providecommand \Eprint [0]{\href }%
\providecommand \doibase [0]{http://dx.doi.org/}%
\providecommand \selectlanguage [0]{\@gobble}%
\providecommand \bibinfo  [0]{\@secondoftwo}%
\providecommand \bibfield  [0]{\@secondoftwo}%
\providecommand \translation [1]{[#1]}%
\providecommand \BibitemOpen [0]{}%
\providecommand \bibitemStop [0]{}%
\providecommand \bibitemNoStop [0]{.\EOS\space}%
\providecommand \EOS [0]{\spacefactor3000\relax}%
\providecommand \BibitemShut  [1]{\csname bibitem#1\endcsname}%
\let\auto@bib@innerbib\@empty
\bibitem [{\citenamefont {Kimble}(2008)}]{Kimble2008a}%
  \BibitemOpen
  \bibfield  {author} {\bibinfo {author} {\bibfnamefont {H.~J.}\ \bibnamefont
  {Kimble}},\ }\href {\doibase 10.1038/nature07127} {\bibfield  {journal}
  {\bibinfo  {journal} {Nature}\ }\textbf {\bibinfo {volume} {453}},\ \bibinfo
  {pages} {1023} (\bibinfo {year} {2008})}\BibitemShut {NoStop}%
\bibitem [{\citenamefont {Gao}\ \emph {et~al.}(2015)\citenamefont {Gao},
  \citenamefont {Imamoglu}, \citenamefont {Bernien},\ and\ \citenamefont
  {Hanson}}]{Gao2015a}%
  \BibitemOpen
  \bibfield  {author} {\bibinfo {author} {\bibfnamefont {W.~B.}\ \bibnamefont
  {Gao}}, \bibinfo {author} {\bibfnamefont {A.}~\bibnamefont {Imamoglu}},
  \bibinfo {author} {\bibfnamefont {H.}~\bibnamefont {Bernien}}, \ and\
  \bibinfo {author} {\bibfnamefont {R.}~\bibnamefont {Hanson}},\ }\href
  {\doibase 10.1038/nphoton.2015.58} {\bibfield  {journal} {\bibinfo  {journal}
  {Nat. Photonics}\ }\textbf {\bibinfo {volume} {9}},\ \bibinfo {pages} {363}
  (\bibinfo {year} {2015})}\BibitemShut {NoStop}%
\bibitem [{\citenamefont {Smith}\ \emph {et~al.}(2005)\citenamefont {Smith},
  \citenamefont {Dalgarno}, \citenamefont {Warburton}, \citenamefont {Govorov},
  \citenamefont {Karrai}, \citenamefont {Gerardot},\ and\ \citenamefont
  {Petroff}}]{Smith2005}%
  \BibitemOpen
  \bibfield  {author} {\bibinfo {author} {\bibfnamefont {J.~M.}\ \bibnamefont
  {Smith}}, \bibinfo {author} {\bibfnamefont {P.~A.}\ \bibnamefont {Dalgarno}},
  \bibinfo {author} {\bibfnamefont {R.~J.}\ \bibnamefont {Warburton}}, \bibinfo
  {author} {\bibfnamefont {A.~O.}\ \bibnamefont {Govorov}}, \bibinfo {author}
  {\bibfnamefont {K.}~\bibnamefont {Karrai}}, \bibinfo {author} {\bibfnamefont
  {B.~D.}\ \bibnamefont {Gerardot}}, \ and\ \bibinfo {author} {\bibfnamefont
  {P.~M.}\ \bibnamefont {Petroff}},\ }\href {\doibase
  10.1103/PhysRevLett.94.197402} {\bibfield  {journal} {\bibinfo  {journal}
  {Phys. Rev. Lett.}\ }\textbf {\bibinfo {volume} {94}},\ \bibinfo {pages}
  {197402} (\bibinfo {year} {2005})}\BibitemShut {NoStop}%
\bibitem [{\citenamefont {Ding}\ \emph {et~al.}(2016)\citenamefont {Ding},
  \citenamefont {He}, \citenamefont {Duan}, \citenamefont {Gregersen},
  \citenamefont {Chen}, \citenamefont {Unsleber}, \citenamefont {Maier},
  \citenamefont {Schneider}, \citenamefont {Kamp}, \citenamefont
  {H{\"{o}}fling}, \citenamefont {Lu},\ and\ \citenamefont {Pan}}]{Ding2016}%
  \BibitemOpen
  \bibfield  {author} {\bibinfo {author} {\bibfnamefont {X.}~\bibnamefont
  {Ding}}, \bibinfo {author} {\bibfnamefont {Y.}~\bibnamefont {He}}, \bibinfo
  {author} {\bibfnamefont {Z.-C.}\ \bibnamefont {Duan}}, \bibinfo {author}
  {\bibfnamefont {N.}~\bibnamefont {Gregersen}}, \bibinfo {author}
  {\bibfnamefont {M.-C.}\ \bibnamefont {Chen}}, \bibinfo {author}
  {\bibfnamefont {S.}~\bibnamefont {Unsleber}}, \bibinfo {author}
  {\bibfnamefont {S.}~\bibnamefont {Maier}}, \bibinfo {author} {\bibfnamefont
  {C.}~\bibnamefont {Schneider}}, \bibinfo {author} {\bibfnamefont
  {M.}~\bibnamefont {Kamp}}, \bibinfo {author} {\bibfnamefont {S.}~\bibnamefont
  {H{\"{o}}fling}}, \bibinfo {author} {\bibfnamefont {C.-Y.}\ \bibnamefont
  {Lu}}, \ and\ \bibinfo {author} {\bibfnamefont {J.-W.}\ \bibnamefont {Pan}},\
  }\href {\doibase 10.1103/PhysRevLett.116.020401} {\bibfield  {journal}
  {\bibinfo  {journal} {Phys. Rev. Lett.}\ }\textbf {\bibinfo {volume} {116}},\
  \bibinfo {pages} {020401} (\bibinfo {year} {2016})}\BibitemShut {NoStop}%
\bibitem [{\citenamefont {Press}\ \emph {et~al.}(2008)\citenamefont {Press},
  \citenamefont {Ladd}, \citenamefont {Zhang},\ and\ \citenamefont
  {Yamamoto}}]{Press2008b}%
  \BibitemOpen
  \bibfield  {author} {\bibinfo {author} {\bibfnamefont {D.}~\bibnamefont
  {Press}}, \bibinfo {author} {\bibfnamefont {T.~D.}\ \bibnamefont {Ladd}},
  \bibinfo {author} {\bibfnamefont {B.}~\bibnamefont {Zhang}}, \ and\ \bibinfo
  {author} {\bibfnamefont {Y.}~\bibnamefont {Yamamoto}},\ }\href {\doibase
  10.1038/nature07530} {\bibfield  {journal} {\bibinfo  {journal} {Nature}\
  }\textbf {\bibinfo {volume} {456}},\ \bibinfo {pages} {218} (\bibinfo {year}
  {2008})}\BibitemShut {NoStop}%
\bibitem [{\citenamefont {{De Greve}}\ \emph {et~al.}(2011)\citenamefont {{De
  Greve}}, \citenamefont {McMahon}, \citenamefont {Press}, \citenamefont
  {Ladd}, \citenamefont {Bisping}, \citenamefont {Schneider}, \citenamefont
  {Kamp}, \citenamefont {Worschech}, \citenamefont {H{\"{o}}fling},
  \citenamefont {Forchel},\ and\ \citenamefont {Yamamoto}}]{DeGreve2011}%
  \BibitemOpen
  \bibfield  {author} {\bibinfo {author} {\bibfnamefont {K.}~\bibnamefont {{De
  Greve}}}, \bibinfo {author} {\bibfnamefont {P.~L.}\ \bibnamefont {McMahon}},
  \bibinfo {author} {\bibfnamefont {D.}~\bibnamefont {Press}}, \bibinfo
  {author} {\bibfnamefont {T.~D.}\ \bibnamefont {Ladd}}, \bibinfo {author}
  {\bibfnamefont {D.}~\bibnamefont {Bisping}}, \bibinfo {author} {\bibfnamefont
  {C.}~\bibnamefont {Schneider}}, \bibinfo {author} {\bibfnamefont
  {M.}~\bibnamefont {Kamp}}, \bibinfo {author} {\bibfnamefont {L.}~\bibnamefont
  {Worschech}}, \bibinfo {author} {\bibfnamefont {S.}~\bibnamefont
  {H{\"{o}}fling}}, \bibinfo {author} {\bibfnamefont {A.}~\bibnamefont
  {Forchel}}, \ and\ \bibinfo {author} {\bibfnamefont {Y.}~\bibnamefont
  {Yamamoto}},\ }\href {\doibase 10.1038/nphys2078} {\bibfield  {journal}
  {\bibinfo  {journal} {Nat. Phys.}\ }\textbf {\bibinfo {volume} {7}},\
  \bibinfo {pages} {872} (\bibinfo {year} {2011})}\BibitemShut {NoStop}%
\bibitem [{\citenamefont {{De Greve}}\ \emph {et~al.}(2012)\citenamefont {{De
  Greve}}, \citenamefont {Yu}, \citenamefont {McMahon}, \citenamefont {Pelc},
  \citenamefont {Natarajan}, \citenamefont {Kim}, \citenamefont {Abe},
  \citenamefont {Maier}, \citenamefont {Schneider}, \citenamefont {Kamp},
  \citenamefont {H{\"{o}}fling}, \citenamefont {Hadfield}, \citenamefont
  {Forchel}, \citenamefont {Fejer},\ and\ \citenamefont
  {Yamamoto}}]{DeGreve2012b}%
  \BibitemOpen
  \bibfield  {author} {\bibinfo {author} {\bibfnamefont {K.}~\bibnamefont {{De
  Greve}}}, \bibinfo {author} {\bibfnamefont {L.}~\bibnamefont {Yu}}, \bibinfo
  {author} {\bibfnamefont {P.~L.}\ \bibnamefont {McMahon}}, \bibinfo {author}
  {\bibfnamefont {J.~S.}\ \bibnamefont {Pelc}}, \bibinfo {author}
  {\bibfnamefont {C.~M.}\ \bibnamefont {Natarajan}}, \bibinfo {author}
  {\bibfnamefont {N.~Y.}\ \bibnamefont {Kim}}, \bibinfo {author} {\bibfnamefont
  {E.}~\bibnamefont {Abe}}, \bibinfo {author} {\bibfnamefont {S.}~\bibnamefont
  {Maier}}, \bibinfo {author} {\bibfnamefont {C.}~\bibnamefont {Schneider}},
  \bibinfo {author} {\bibfnamefont {M.}~\bibnamefont {Kamp}}, \bibinfo {author}
  {\bibfnamefont {S.}~\bibnamefont {H{\"{o}}fling}}, \bibinfo {author}
  {\bibfnamefont {R.~H.}\ \bibnamefont {Hadfield}}, \bibinfo {author}
  {\bibfnamefont {A.}~\bibnamefont {Forchel}}, \bibinfo {author} {\bibfnamefont
  {M.~M.}\ \bibnamefont {Fejer}}, \ and\ \bibinfo {author} {\bibfnamefont
  {Y.}~\bibnamefont {Yamamoto}},\ }\href {\doibase 10.1038/nature11577}
  {\bibfield  {journal} {\bibinfo  {journal} {Nature}\ }\textbf {\bibinfo
  {volume} {491}},\ \bibinfo {pages} {421} (\bibinfo {year}
  {2012})}\BibitemShut {NoStop}%
\bibitem [{\citenamefont {Gao}\ \emph {et~al.}(2012)\citenamefont {Gao},
  \citenamefont {Fallahi}, \citenamefont {Togan}, \citenamefont
  {Miguel-Sanchez},\ and\ \citenamefont {Imamoglu}}]{Gao2012a}%
  \BibitemOpen
  \bibfield  {author} {\bibinfo {author} {\bibfnamefont {W.~B.}\ \bibnamefont
  {Gao}}, \bibinfo {author} {\bibfnamefont {P.}~\bibnamefont {Fallahi}},
  \bibinfo {author} {\bibfnamefont {E.}~\bibnamefont {Togan}}, \bibinfo
  {author} {\bibfnamefont {J.}~\bibnamefont {Miguel-Sanchez}}, \ and\ \bibinfo
  {author} {\bibfnamefont {A.}~\bibnamefont {Imamoglu}},\ }\href {\doibase
  10.1038/nature11573} {\bibfield  {journal} {\bibinfo  {journal} {Nature}\
  }\textbf {\bibinfo {volume} {491}},\ \bibinfo {pages} {426} (\bibinfo {year}
  {2012})}\BibitemShut {NoStop}%
\bibitem [{\citenamefont {Schaibley}\ \emph {et~al.}(2013)\citenamefont
  {Schaibley}, \citenamefont {Burgers}, \citenamefont {McCracken},
  \citenamefont {Duan}, \citenamefont {Berman}, \citenamefont {Steel},
  \citenamefont {Bracker}, \citenamefont {Gammon},\ and\ \citenamefont
  {Sham}}]{Schaibley2013}%
  \BibitemOpen
  \bibfield  {author} {\bibinfo {author} {\bibfnamefont {J.~R.}\ \bibnamefont
  {Schaibley}}, \bibinfo {author} {\bibfnamefont {A.~P.}\ \bibnamefont
  {Burgers}}, \bibinfo {author} {\bibfnamefont {G.~A.}\ \bibnamefont
  {McCracken}}, \bibinfo {author} {\bibfnamefont {L.-M.}\ \bibnamefont {Duan}},
  \bibinfo {author} {\bibfnamefont {P.~R.}\ \bibnamefont {Berman}}, \bibinfo
  {author} {\bibfnamefont {D.~G.}\ \bibnamefont {Steel}}, \bibinfo {author}
  {\bibfnamefont {A.~S.}\ \bibnamefont {Bracker}}, \bibinfo {author}
  {\bibfnamefont {D.}~\bibnamefont {Gammon}}, \ and\ \bibinfo {author}
  {\bibfnamefont {L.~J.}\ \bibnamefont {Sham}},\ }\href {\doibase
  10.1103/PhysRevLett.110.167401} {\bibfield  {journal} {\bibinfo  {journal}
  {Phys. Rev. Lett.}\ }\textbf {\bibinfo {volume} {110}},\ \bibinfo {pages}
  {167401} (\bibinfo {year} {2013})}\BibitemShut {NoStop}%
\bibitem [{\citenamefont {Delteil}\ \emph {et~al.}(2015)\citenamefont
  {Delteil}, \citenamefont {Sun}, \citenamefont {Gao}, \citenamefont {Togan},
  \citenamefont {Faelt},\ and\ \citenamefont {Imamoglu}}]{Delteil2015}%
  \BibitemOpen
  \bibfield  {author} {\bibinfo {author} {\bibfnamefont {A.}~\bibnamefont
  {Delteil}}, \bibinfo {author} {\bibfnamefont {Z.}~\bibnamefont {Sun}},
  \bibinfo {author} {\bibfnamefont {W.-b.}\ \bibnamefont {Gao}}, \bibinfo
  {author} {\bibfnamefont {E.}~\bibnamefont {Togan}}, \bibinfo {author}
  {\bibfnamefont {S.}~\bibnamefont {Faelt}}, \ and\ \bibinfo {author}
  {\bibfnamefont {A.}~\bibnamefont {Imamoglu}},\ }\href {\doibase
  10.1038/nphys3605} {\bibfield  {journal} {\bibinfo  {journal} {Nat. Phys.}\
  }\textbf {\bibinfo {volume} {12}},\ \bibinfo {pages} {218} (\bibinfo {year}
  {2015})}\BibitemShut {NoStop}%
\bibitem [{\citenamefont {Stockill}\ \emph {et~al.}(2017)\citenamefont
  {Stockill}, \citenamefont {Stanley}, \citenamefont {Huthmacher},
  \citenamefont {Clarke}, \citenamefont {Hugues}, \citenamefont {Miller},
  \citenamefont {Matthiesen}, \citenamefont {{Le Gall}},\ and\ \citenamefont
  {Atat{\"{u}}re}}]{Stockill2017}%
  \BibitemOpen
  \bibfield  {author} {\bibinfo {author} {\bibfnamefont {R.}~\bibnamefont
  {Stockill}}, \bibinfo {author} {\bibfnamefont {M.~J.}\ \bibnamefont
  {Stanley}}, \bibinfo {author} {\bibfnamefont {L.}~\bibnamefont {Huthmacher}},
  \bibinfo {author} {\bibfnamefont {E.}~\bibnamefont {Clarke}}, \bibinfo
  {author} {\bibfnamefont {M.}~\bibnamefont {Hugues}}, \bibinfo {author}
  {\bibfnamefont {A.~J.}\ \bibnamefont {Miller}}, \bibinfo {author}
  {\bibfnamefont {C.}~\bibnamefont {Matthiesen}}, \bibinfo {author}
  {\bibfnamefont {C.}~\bibnamefont {{Le Gall}}}, \ and\ \bibinfo {author}
  {\bibfnamefont {M.}~\bibnamefont {Atat{\"{u}}re}},\ }\href {\doibase
  10.1103/PhysRevLett.119.010503} {\bibfield  {journal} {\bibinfo  {journal}
  {Phys. Rev. Lett.}\ }\textbf {\bibinfo {volume} {119}},\ \bibinfo {pages}
  {010503} (\bibinfo {year} {2017})}\BibitemShut {NoStop}%
\bibitem [{\citenamefont {Bechtold}\ \emph {et~al.}(2015)\citenamefont
  {Bechtold}, \citenamefont {Rauch}, \citenamefont {Li}, \citenamefont
  {Simmet}, \citenamefont {Ardelt}, \citenamefont {Regler}, \citenamefont
  {M{\"{u}}ller}, \citenamefont {Sinitsyn},\ and\ \citenamefont
  {Finley}}]{Bechtold2015a}%
  \BibitemOpen
  \bibfield  {author} {\bibinfo {author} {\bibfnamefont {A.}~\bibnamefont
  {Bechtold}}, \bibinfo {author} {\bibfnamefont {D.}~\bibnamefont {Rauch}},
  \bibinfo {author} {\bibfnamefont {F.}~\bibnamefont {Li}}, \bibinfo {author}
  {\bibfnamefont {T.}~\bibnamefont {Simmet}}, \bibinfo {author} {\bibfnamefont
  {P.-L.}\ \bibnamefont {Ardelt}}, \bibinfo {author} {\bibfnamefont
  {A.}~\bibnamefont {Regler}}, \bibinfo {author} {\bibfnamefont
  {K.}~\bibnamefont {M{\"{u}}ller}}, \bibinfo {author} {\bibfnamefont {N.~A.}\
  \bibnamefont {Sinitsyn}}, \ and\ \bibinfo {author} {\bibfnamefont {J.~J.}\
  \bibnamefont {Finley}},\ }\href {\doibase 10.1038/nphys3470} {\bibfield
  {journal} {\bibinfo  {journal} {Nat. Phys.}\ }\textbf {\bibinfo {volume}
  {11}},\ \bibinfo {pages} {1005} (\bibinfo {year} {2015})}\BibitemShut
  {NoStop}%
\bibitem [{\citenamefont {Stockill}\ \emph {et~al.}(2016)\citenamefont
  {Stockill}, \citenamefont {{Le Gall}}, \citenamefont {Matthiesen},
  \citenamefont {Huthmacher}, \citenamefont {Clarke}, \citenamefont {Hugues},\
  and\ \citenamefont {Atat{\"{u}}re}}]{Stockill2016}%
  \BibitemOpen
  \bibfield  {author} {\bibinfo {author} {\bibfnamefont {R.}~\bibnamefont
  {Stockill}}, \bibinfo {author} {\bibfnamefont {C.}~\bibnamefont {{Le Gall}}},
  \bibinfo {author} {\bibfnamefont {C.}~\bibnamefont {Matthiesen}}, \bibinfo
  {author} {\bibfnamefont {L.}~\bibnamefont {Huthmacher}}, \bibinfo {author}
  {\bibfnamefont {E.}~\bibnamefont {Clarke}}, \bibinfo {author} {\bibfnamefont
  {M.}~\bibnamefont {Hugues}}, \ and\ \bibinfo {author} {\bibfnamefont
  {M.}~\bibnamefont {Atat{\"{u}}re}},\ }\href {\doibase 10.1038/ncomms12745}
  {\bibfield  {journal} {\bibinfo  {journal} {Nat. Commun.}\ }\textbf {\bibinfo
  {volume} {7}},\ \bibinfo {pages} {12745} (\bibinfo {year}
  {2016})}\BibitemShut {NoStop}%
\bibitem [{\citenamefont {Fischer}\ \emph {et~al.}(2008)\citenamefont
  {Fischer}, \citenamefont {Coish}, \citenamefont {Bulaev},\ and\ \citenamefont
  {Loss}}]{Fischer2008}%
  \BibitemOpen
  \bibfield  {author} {\bibinfo {author} {\bibfnamefont {J.}~\bibnamefont
  {Fischer}}, \bibinfo {author} {\bibfnamefont {W.~A.}\ \bibnamefont {Coish}},
  \bibinfo {author} {\bibfnamefont {D.~V.}\ \bibnamefont {Bulaev}}, \ and\
  \bibinfo {author} {\bibfnamefont {D.}~\bibnamefont {Loss}},\ }\href {\doibase
  10.1103/PhysRevB.78.155329} {\bibfield  {journal} {\bibinfo  {journal} {Phys.
  Rev. B}\ }\textbf {\bibinfo {volume} {78}},\ \bibinfo {pages} {155329}
  (\bibinfo {year} {2008})}\BibitemShut {NoStop}%
\bibitem [{\citenamefont {Fallahi}\ \emph {et~al.}(2010)\citenamefont
  {Fallahi}, \citenamefont {Yilmaz},\ and\ \citenamefont
  {Imamoglu}}]{Fallahi2010}%
  \BibitemOpen
  \bibfield  {author} {\bibinfo {author} {\bibfnamefont {P.}~\bibnamefont
  {Fallahi}}, \bibinfo {author} {\bibfnamefont {S.~T.}\ \bibnamefont
  {Yilmaz}}, \ and\ \bibinfo {author} {\bibfnamefont {A.}~\bibnamefont
  {Imamoglu}},\ }\href {\doibase 10.1103/PhysRevLett.105.257402} {\bibfield
  {journal} {\bibinfo  {journal} {Phys. Rev. Lett.}\ }\textbf {\bibinfo
  {volume} {105}},\ \bibinfo {pages} {257402} (\bibinfo {year}
  {2010})}\BibitemShut {NoStop}%
\bibitem [{\citenamefont {Houel}\ \emph {et~al.}(2014)\citenamefont {Houel},
  \citenamefont {Prechtel}, \citenamefont {Kuhlmann}, \citenamefont {Brunner},
  \citenamefont {Kuklewicz}, \citenamefont {Gerardot}, \citenamefont {Stoltz},
  \citenamefont {Petroff},\ and\ \citenamefont {Warburton}}]{Houel2014}%
  \BibitemOpen
  \bibfield  {author} {\bibinfo {author} {\bibfnamefont {J.}~\bibnamefont
  {Houel}}, \bibinfo {author} {\bibfnamefont {J.~H.}\ \bibnamefont {Prechtel}},
  \bibinfo {author} {\bibfnamefont {A.~V.}\ \bibnamefont {Kuhlmann}}, \bibinfo
  {author} {\bibfnamefont {D.}~\bibnamefont {Brunner}}, \bibinfo {author}
  {\bibfnamefont {C.~E.}\ \bibnamefont {Kuklewicz}}, \bibinfo {author}
  {\bibfnamefont {B.~D.}\ \bibnamefont {Gerardot}}, \bibinfo {author}
  {\bibfnamefont {N.~G.}\ \bibnamefont {Stoltz}}, \bibinfo {author}
  {\bibfnamefont {P.~M.}\ \bibnamefont {Petroff}}, \ and\ \bibinfo {author}
  {\bibfnamefont {R.~J.}\ \bibnamefont {Warburton}},\ }\href {\doibase
  10.1103/PhysRevLett.112.107401} {\bibfield  {journal} {\bibinfo  {journal}
  {Phys. Rev. Lett.}\ }\textbf {\bibinfo {volume} {112}},\ \bibinfo {pages}
  {107401} (\bibinfo {year} {2014})}\BibitemShut {NoStop}%
\bibitem [{\citenamefont {Prechtel}\ \emph {et~al.}(2016)\citenamefont
  {Prechtel}, \citenamefont {Kuhlmann}, \citenamefont {Houel}, \citenamefont
  {Ludwig}, \citenamefont {Valentin}, \citenamefont {Wieck},\ and\
  \citenamefont {Warburton}}]{Prechtel2016}%
  \BibitemOpen
  \bibfield  {author} {\bibinfo {author} {\bibfnamefont {J.~H.}\ \bibnamefont
  {Prechtel}}, \bibinfo {author} {\bibfnamefont {A.~V.}\ \bibnamefont
  {Kuhlmann}}, \bibinfo {author} {\bibfnamefont {J.}~\bibnamefont {Houel}},
  \bibinfo {author} {\bibfnamefont {A.}~\bibnamefont {Ludwig}}, \bibinfo
  {author} {\bibfnamefont {S.~R.}\ \bibnamefont {Valentin}}, \bibinfo {author}
  {\bibfnamefont {A.~D.}\ \bibnamefont {Wieck}}, \ and\ \bibinfo {author}
  {\bibfnamefont {R.~J.}\ \bibnamefont {Warburton}},\ }\href {\doibase
  10.1038/nmat4704} {\bibfield  {journal} {\bibinfo  {journal} {Nat. Mater.}\
  }\textbf {\bibinfo {volume} {15}},\ \bibinfo {pages} {981} (\bibinfo {year}
  {2016})}\BibitemShut {NoStop}%
\bibitem [{\citenamefont {Greilich}\ \emph {et~al.}(2011)\citenamefont
  {Greilich}, \citenamefont {Carter}, \citenamefont {Kim}, \citenamefont
  {Bracker},\ and\ \citenamefont {Gammon}}]{Greilich2011}%
  \BibitemOpen
  \bibfield  {author} {\bibinfo {author} {\bibfnamefont {A.}~\bibnamefont
  {Greilich}}, \bibinfo {author} {\bibfnamefont {S.~G.}\ \bibnamefont
  {Carter}}, \bibinfo {author} {\bibfnamefont {D.}~\bibnamefont {Kim}},
  \bibinfo {author} {\bibfnamefont {A.~S.}\ \bibnamefont {Bracker}}, \ and\
  \bibinfo {author} {\bibfnamefont {D.}~\bibnamefont {Gammon}},\ }\href
  {\doibase 10.1038/nphoton.2011.237} {\bibfield  {journal} {\bibinfo
  {journal} {Nat. Photonics}\ }\textbf {\bibinfo {volume} {5}},\ \bibinfo
  {pages} {702} (\bibinfo {year} {2011})}\BibitemShut {NoStop}%
\bibitem [{\citenamefont {Godden}\ \emph {et~al.}(2012)\citenamefont {Godden},
  \citenamefont {Quilter}, \citenamefont {Ramsay}, \citenamefont {Wu},
  \citenamefont {Brereton}, \citenamefont {Boyle}, \citenamefont {Luxmoore},
  \citenamefont {Puebla-Nunez}, \citenamefont {Fox},\ and\ \citenamefont
  {Skolnick}}]{Godden2012}%
  \BibitemOpen
  \bibfield  {author} {\bibinfo {author} {\bibfnamefont {T.~M.}\ \bibnamefont
  {Godden}}, \bibinfo {author} {\bibfnamefont {J.~H.}\ \bibnamefont {Quilter}},
  \bibinfo {author} {\bibfnamefont {A.~J.}\ \bibnamefont {Ramsay}}, \bibinfo
  {author} {\bibfnamefont {Y.}~\bibnamefont {Wu}}, \bibinfo {author}
  {\bibfnamefont {P.}~\bibnamefont {Brereton}}, \bibinfo {author}
  {\bibfnamefont {S.~J.}\ \bibnamefont {Boyle}}, \bibinfo {author}
  {\bibfnamefont {I.~J.}\ \bibnamefont {Luxmoore}}, \bibinfo {author}
  {\bibfnamefont {J.}~\bibnamefont {Puebla-Nunez}}, \bibinfo {author}
  {\bibfnamefont {A.~M.}\ \bibnamefont {Fox}}, \ and\ \bibinfo {author}
  {\bibfnamefont {M.~S.}\ \bibnamefont {Skolnick}},\ }\href {\doibase
  10.1103/PhysRevLett.108.017402} {\bibfield  {journal} {\bibinfo  {journal}
  {Phys. Rev. Lett.}\ }\textbf {\bibinfo {volume} {108}},\ \bibinfo {pages}
  {017402} (\bibinfo {year} {2012})}\BibitemShut {NoStop}%
\bibitem [{\citenamefont {Sun}\ \emph {et~al.}(2016)\citenamefont {Sun},
  \citenamefont {Delteil}, \citenamefont {Faelt},\ and\ \citenamefont
  {Imamoglu}}]{Sun2016}%
  \BibitemOpen
  \bibfield  {author} {\bibinfo {author} {\bibfnamefont {Z.}~\bibnamefont
  {Sun}}, \bibinfo {author} {\bibfnamefont {A.}~\bibnamefont {Delteil}},
  \bibinfo {author} {\bibfnamefont {S.}~\bibnamefont {Faelt}}, \ and\ \bibinfo
  {author} {\bibfnamefont {A.}~\bibnamefont {Imamoglu}},\ }\href {\doibase
  10.1103/PhysRevB.93.241302} {\bibfield  {journal} {\bibinfo  {journal} {Phys.
  Rev. B}\ }\textbf {\bibinfo {volume} {93}},\ \bibinfo {pages} {241302}
  (\bibinfo {year} {2016})}\BibitemShut {NoStop}%
\bibitem [{Sup()}]{Supp}%
  \BibitemOpen
  \href@noop {} {}\bibinfo {note} {See Supplemental Material}\BibitemShut {Stop}%
\bibitem [{\citenamefont {Salis}\ \emph {et~al.}(2001)\citenamefont {Salis},
  \citenamefont {Kato}, \citenamefont {Ensslin}, \citenamefont {Driscoll},
  \citenamefont {Gossard},\ and\ \citenamefont {Awschalom}}]{Salis2001}%
  \BibitemOpen
  \bibfield  {author} {\bibinfo {author} {\bibfnamefont {G.}~\bibnamefont
  {Salis}}, \bibinfo {author} {\bibfnamefont {Y.}~\bibnamefont {Kato}},
  \bibinfo {author} {\bibfnamefont {K.}~\bibnamefont {Ensslin}}, \bibinfo
  {author} {\bibfnamefont {D.~C.}\ \bibnamefont {Driscoll}}, \bibinfo {author}
  {\bibfnamefont {A.~C.}\ \bibnamefont {Gossard}}, \ and\ \bibinfo {author}
  {\bibfnamefont {D.~D.}\ \bibnamefont {Awschalom}},\ }\href {\doibase
  10.1038/414619a} {\bibfield  {journal} {\bibinfo  {journal} {Nature}\
  }\textbf {\bibinfo {volume} {414}},\ \bibinfo {pages} {619} (\bibinfo {year}
  {2001})}\BibitemShut {NoStop}%
\bibitem [{\citenamefont {Testelin}\ \emph {et~al.}(2009)\citenamefont
  {Testelin}, \citenamefont {Bernardot}, \citenamefont {Eble},\ and\
  \citenamefont {Chamarro}}]{Testelin2009}%
  \BibitemOpen
  \bibfield  {author} {\bibinfo {author} {\bibfnamefont {C.}~\bibnamefont
  {Testelin}}, \bibinfo {author} {\bibfnamefont {F.}~\bibnamefont {Bernardot}},
  \bibinfo {author} {\bibfnamefont {B.}~\bibnamefont {Eble}}, \ and\ \bibinfo
  {author} {\bibfnamefont {M.}~\bibnamefont {Chamarro}},\ }\href {\doibase
  10.1103/PhysRevB.79.195440} {\bibfield  {journal} {\bibinfo  {journal} {Phys.
  Rev. B}\ }\textbf {\bibinfo {volume} {79}},\ \bibinfo {pages} {195440}
  (\bibinfo {year} {2009})}\BibitemShut {NoStop}%
\bibitem [{\citenamefont {Eble}\ \emph {et~al.}(2009)\citenamefont {Eble},
  \citenamefont {Testelin}, \citenamefont {Desfonds}, \citenamefont
  {Bernardot}, \citenamefont {Balocchi}, \citenamefont {Amand}, \citenamefont
  {Miard}, \citenamefont {Lema{\^{i}}tre}, \citenamefont {Marie},\ and\
  \citenamefont {Chamarro}}]{Eble2009}%
  \BibitemOpen
  \bibfield  {author} {\bibinfo {author} {\bibfnamefont {B.}~\bibnamefont
  {Eble}}, \bibinfo {author} {\bibfnamefont {C.}~\bibnamefont {Testelin}},
  \bibinfo {author} {\bibfnamefont {P.}~\bibnamefont {Desfonds}}, \bibinfo
  {author} {\bibfnamefont {F.}~\bibnamefont {Bernardot}}, \bibinfo {author}
  {\bibfnamefont {A.}~\bibnamefont {Balocchi}}, \bibinfo {author}
  {\bibfnamefont {T.}~\bibnamefont {Amand}}, \bibinfo {author} {\bibfnamefont
  {A.}~\bibnamefont {Miard}}, \bibinfo {author} {\bibfnamefont
  {A.}~\bibnamefont {Lema{\^{i}}tre}}, \bibinfo {author} {\bibfnamefont
  {X.}~\bibnamefont {Marie}}, \ and\ \bibinfo {author} {\bibfnamefont
  {M.}~\bibnamefont {Chamarro}},\ }\href {\doibase
  10.1103/PhysRevLett.102.146601} {\bibfield  {journal} {\bibinfo  {journal}
  {Phys. Rev. Lett.}\ }\textbf {\bibinfo {volume} {102}},\ \bibinfo {pages}
  {146601} (\bibinfo {year} {2009})}\BibitemShut {NoStop}%
\bibitem [{\citenamefont {Dahbashi}\ \emph {et~al.}(2012)\citenamefont
  {Dahbashi}, \citenamefont {H{\"{u}}bner}, \citenamefont {Berski},
  \citenamefont {Wiegand}, \citenamefont {Marie}, \citenamefont {Pierz},
  \citenamefont {Schumacher},\ and\ \citenamefont {Oestreich}}]{Dahbashi2012}%
  \BibitemOpen
  \bibfield  {author} {\bibinfo {author} {\bibfnamefont {R.}~\bibnamefont
  {Dahbashi}}, \bibinfo {author} {\bibfnamefont {J.}~\bibnamefont
  {H{\"{u}}bner}}, \bibinfo {author} {\bibfnamefont {F.}~\bibnamefont
  {Berski}}, \bibinfo {author} {\bibfnamefont {J.}~\bibnamefont {Wiegand}},
  \bibinfo {author} {\bibfnamefont {X.}~\bibnamefont {Marie}}, \bibinfo
  {author} {\bibfnamefont {K.}~\bibnamefont {Pierz}}, \bibinfo {author}
  {\bibfnamefont {H.~W.}\ \bibnamefont {Schumacher}}, \ and\ \bibinfo {author}
  {\bibfnamefont {M.}~\bibnamefont {Oestreich}},\ }\href {\doibase
  10.1063/1.3678182} {\bibfield  {journal} {\bibinfo  {journal} {Appl. Phys.
  Lett.}\ }\textbf {\bibinfo {volume} {100}},\ \bibinfo {pages} {031906}
  (\bibinfo {year} {2012})}\BibitemShut {NoStop}%
\bibitem [{\citenamefont {Hahn}(1950)}]{Hahn1950}%
  \BibitemOpen
  \bibfield  {author} {\bibinfo {author} {\bibfnamefont {E.~L.}\ \bibnamefont
  {Hahn}},\ }\href {\doibase 10.1103/PhysRev.80.580} {\bibfield  {journal}
  {\bibinfo  {journal} {Phys. Rev.}\ }\textbf {\bibinfo {volume} {80}},\
  \bibinfo {pages} {580} (\bibinfo {year} {1950})}\BibitemShut {NoStop}%
\bibitem [{Note1()}]{Note1}%
  \BibitemOpen
  \bibinfo {note} {The choice of $\alpha $ is motivated by the scaling observed
  in dynamic decoupling \cite {Supp}}\BibitemShut {NoStop}%
\bibitem [{\citenamefont {Kuhlmann}\ \emph {et~al.}(2013)\citenamefont
  {Kuhlmann}, \citenamefont {Houel}, \citenamefont {Ludwig}, \citenamefont
  {Greuter}, \citenamefont {Reuter}, \citenamefont {Wieck}, \citenamefont
  {Poggio},\ and\ \citenamefont {Warburton}}]{Kuhlmann2013a}%
  \BibitemOpen
  \bibfield  {author} {\bibinfo {author} {\bibfnamefont {A.~V.}\ \bibnamefont
  {Kuhlmann}}, \bibinfo {author} {\bibfnamefont {J.}~\bibnamefont {Houel}},
  \bibinfo {author} {\bibfnamefont {A.}~\bibnamefont {Ludwig}}, \bibinfo
  {author} {\bibfnamefont {L.}~\bibnamefont {Greuter}}, \bibinfo {author}
  {\bibfnamefont {D.}~\bibnamefont {Reuter}}, \bibinfo {author} {\bibfnamefont
  {A.~D.}\ \bibnamefont {Wieck}}, \bibinfo {author} {\bibfnamefont
  {M.}~\bibnamefont {Poggio}}, \ and\ \bibinfo {author} {\bibfnamefont {R.~J.}\
  \bibnamefont {Warburton}},\ }\href {\doibase 10.1038/nphys2688} {\bibfield
  {journal} {\bibinfo  {journal} {Nat. Phys.}\ }\textbf {\bibinfo {volume}
  {9}},\ \bibinfo {pages} {570} (\bibinfo {year} {2013})}\BibitemShut {NoStop}%
\bibitem [{\citenamefont {Stanley}\ \emph {et~al.}(2014)\citenamefont
  {Stanley}, \citenamefont {Matthiesen}, \citenamefont {Hansom}, \citenamefont
  {{Le Gall}}, \citenamefont {Schulte}, \citenamefont {Clarke},\ and\
  \citenamefont {Atat{\"{u}}re}}]{Stanley2014}%
  \BibitemOpen
  \bibfield  {author} {\bibinfo {author} {\bibfnamefont {M.~J.}\ \bibnamefont
  {Stanley}}, \bibinfo {author} {\bibfnamefont {C.}~\bibnamefont {Matthiesen}},
  \bibinfo {author} {\bibfnamefont {J.}~\bibnamefont {Hansom}}, \bibinfo
  {author} {\bibfnamefont {C.}~\bibnamefont {{Le Gall}}}, \bibinfo {author}
  {\bibfnamefont {C.~H.~H.}\ \bibnamefont {Schulte}}, \bibinfo {author}
  {\bibfnamefont {E.}~\bibnamefont {Clarke}}, \ and\ \bibinfo {author}
  {\bibfnamefont {M.}~\bibnamefont {Atat{\"{u}}re}},\ }\href {\doibase
  10.1103/PhysRevB.90.195305} {\bibfield  {journal} {\bibinfo  {journal} {Phys.
  Rev. B}\ }\textbf {\bibinfo {volume} {90}},\ \bibinfo {pages} {195305}
  (\bibinfo {year} {2014})}\BibitemShut {NoStop}%
\bibitem [{\citenamefont {Keshner}(1982)}]{Keshner1982}%
  \BibitemOpen
  \bibfield  {author} {\bibinfo {author} {\bibfnamefont {M.}~\bibnamefont
  {Keshner}},\ }\href {\doibase 10.1109/PROC.1982.12282} {\bibfield  {journal}
  {\bibinfo  {journal} {Proc. IEEE}\ }\textbf {\bibinfo {volume} {70}},\
  \bibinfo {pages} {212} (\bibinfo {year} {1982})}\BibitemShut {NoStop}%
\bibitem [{\citenamefont {Dial}\ \emph {et~al.}(2013)\citenamefont {Dial},
  \citenamefont {Shulman}, \citenamefont {Harvey}, \citenamefont {Bluhm},
  \citenamefont {Umansky},\ and\ \citenamefont {Yacoby}}]{Dial2013}%
  \BibitemOpen
  \bibfield  {author} {\bibinfo {author} {\bibfnamefont {O.~E.}\ \bibnamefont
  {Dial}}, \bibinfo {author} {\bibfnamefont {M.~D.}\ \bibnamefont {Shulman}},
  \bibinfo {author} {\bibfnamefont {S.~P.}\ \bibnamefont {Harvey}}, \bibinfo
  {author} {\bibfnamefont {H.}~\bibnamefont {Bluhm}}, \bibinfo {author}
  {\bibfnamefont {V.}~\bibnamefont {Umansky}}, \ and\ \bibinfo {author}
  {\bibfnamefont {A.}~\bibnamefont {Yacoby}},\ }\href {\doibase
  10.1103/PhysRevLett.110.146804} {\bibfield  {journal} {\bibinfo  {journal}
  {Phys. Rev. Lett.}\ }\textbf {\bibinfo {volume} {110}},\ \bibinfo {pages}
  {146804} (\bibinfo {year} {2013})}\BibitemShut {NoStop}%
\bibitem [{\citenamefont {Carr}\ and\ \citenamefont
  {Purcell}(1954)}]{CarrH.Y.1983}%
  \BibitemOpen
  \bibfield  {author} {\bibinfo {author} {\bibfnamefont {H.~Y.}\ \bibnamefont
  {Carr}}\ and\ \bibinfo {author} {\bibfnamefont {E.~M.}\ \bibnamefont
  {Purcell}},\ }\href {\doibase 10.1103/PhysRev.94.630} {\bibfield  {journal}
  {\bibinfo  {journal} {Phys. Rev.}\ }\textbf {\bibinfo {volume} {94}},\
  \bibinfo {pages} {630} (\bibinfo {year} {1954})}\BibitemShut {NoStop}%
\bibitem [{\citenamefont {Medford}\ \emph {et~al.}(2012)\citenamefont
  {Medford}, \citenamefont {Cywi{\'{n}}ski}, \citenamefont {Barthel},
  \citenamefont {Marcus}, \citenamefont {Hanson},\ and\ \citenamefont
  {Gossard}}]{Medford2012}%
  \BibitemOpen
  \bibfield  {author} {\bibinfo {author} {\bibfnamefont {J.}~\bibnamefont
  {Medford}}, \bibinfo {author} {\bibfnamefont {{\L}.}~\bibnamefont
  {Cywi{\'{n}}ski}}, \bibinfo {author} {\bibfnamefont {C.}~\bibnamefont
  {Barthel}}, \bibinfo {author} {\bibfnamefont {C.~M.}\ \bibnamefont {Marcus}},
  \bibinfo {author} {\bibfnamefont {M.~P.}\ \bibnamefont {Hanson}}, \ and\
  \bibinfo {author} {\bibfnamefont {A.~C.}\ \bibnamefont {Gossard}},\ }\href
  {\doibase 10.1103/PhysRevLett.108.086802} {\bibfield  {journal} {\bibinfo
  {journal} {Phys. Rev. Lett.}\ }\textbf {\bibinfo {volume} {108}},\ \bibinfo
  {pages} {086802} (\bibinfo {year} {2012})}\BibitemShut {NoStop}%
\bibitem [{\citenamefont {Levitt}(1986)}]{Levit1986}%
  \BibitemOpen
  \bibfield  {author} {\bibinfo {author} {\bibfnamefont {M.~H.}\ \bibnamefont
  {Levitt}},\ }\href {\doibase 10.1016/0079-6565(86)80005-X} {\bibfield
  {journal} {\bibinfo  {journal} {Prog. Nucl. Magn. Reson. Spectrosc.}\
  }\textbf {\bibinfo {volume} {18}},\ \bibinfo {pages} {61} (\bibinfo {year}
  {1986})}\BibitemShut {NoStop}%
\end{thebibliography}
\end{document}